\RequirePackage{fix-cm}
\documentclass[twocolumn,epjc3]{svjour3}
\makeatletter
\expandafter\let\csname opt@amsmath.sty\endcsname\relax
\AtBeginDocument{
  \mathindent=15pt 
  \@mathmargin\@centering} 
\makeatother
\usepackage{blindtext}
\usepackage{amsmath,amsfonts,mathtools}
\usepackage{amsfonts}
\usepackage{setspace}
\usepackage[mathletters]{ucs}
\usepackage[utf8x]{inputenc}
\usepackage{url}
\usepackage{graphicx} 
\usepackage{caption}
\usepackage[toc]{appendix}
\usepackage{subcaption}
\usepackage[a4paper, total={7in, 9in}]{geometry}

\journalname{Eur. Phys. J. B}

\begin{document}
\title{Geometric effect on near-field heat transfer analysis using efficient graphene and nanotube models}

\author{Kristo Nugraha Lian\thanksref{addr1,e1} \and Jian-Sheng Wang\thanksref{addr2,e2}
}
\thankstext{e1}{email: kristolian@yahoo.com}
\thankstext{e2}{email: phywjs@nus.edu.sg}
\institute{Centre for Quantum Technologies, National University of Singapore, Singapore 117543, Republic of Singapore \label{addr1}
\and Department of Physics, National University of Singapore, Singapore 117551, Republic of Singapore \label{addr2}
}
\date{Received: / Accepted:}

\maketitle


\begin{abstract}
Following the recent research enthusiasm on the effect of geometry on near-field heat transfer (NFHT) enhancement, we present an analysis based on simplified yet highly efficient graphene and nanotube models.  Two geometries are considered: that of two parallel infinite “graphene" surfaces and that of a one-dimensional infinite “nanotube" line in parallel with an infinite surface.  Due to its symmetry, the former is in principal simpler to analyze and even so, earlier works suggested that the application of a full model in this problem still demands heavy computations.  Among other findings, our simplified computation - having successfully replicated the results of relevant earlier works - suggests a sharper NFHT enhancement dependence on distance for the line-surface system, namely $J\sim d^{-5.1}$ as compared to $J\sim d^{-2.2}$ for the parallel surface.  Such comparisons together with applications of our efficient approach would be the important first steps in the attempt to find a general rule describing geometric dependence of NFHT.
\end{abstract}

\section{Introduction}
In light of recent advancements in nano-materials and design, an amendment to the conventional theory of radiative heat transfer discovered in the 1900s \cite{Planck} is imperative.  Concerning distances of the order of thermal wavelength $\lambda_{th}={\hbar c}/{k_BT}$ or less, electromagnetic waves no longer hold the crucial role as the sole heat transfer mediator; interactions of electrons, plasmons and polaritons begin to gain importance \cite{Jiang,Keller,Mahan0}.  In this regard, research interests grow in the field of near-field radiative heat transfer (NFHT), which was pioneered in the 1970s when Polder and van Hove \cite{PvH} developed the idea of applying the formalism of fluctuational electrodynamics \cite{Rytov} into materials property problem \cite{Main}. Application of this idea on NFHT follows the establishment of the analogue of Poynting vector \cite{Jiebin} in the case of non-photonic heat transfer using the Maxwell's equations.  On the practical spectrum, these theoretical predictions have in fact been realized several years earlier \cite{Hargreaves,Damoto} while new advanced thermal devices, e.g., thermal microscopy (STM), photovoltaic systems, and thermal transistors are unceasingly being developed based on the NFHT principles \cite{Abdallah,Liu}. \\

Following some previous works on NFHT \cite{Jiang,Main}, in this work our analysis considers only the contribution of charge fluctuations and their corresponding “scalar photons" from the scalar field.  This point has indeed been discussed by some previous works, for example Keller \cite{Keller} on the neglection of the propagating field terms and Abajo \cite{Abajo} on the neglection of the plasmon's retardation factor, the latter being stemmed from the fact that graphene's plasmon wavelengths are typically a few orders smaller than their thermal counterpart.  \\

For the purpose of our study, we are primarily interested in the global geometry of the system; this motivates our simplifications of the local graphene and nanotube structures. Our objective is to determine the asymptotic effective exponent of the heat transfer vs distance curve, in which case the conclusion shall no longer be significantly affected by such less-than-nanoscale variation.  Furthermore, comparisons to previous works such as Jiang and Wang \cite{Jiang} shall suffice to prove this point.  In fact, such comparisons would facilitate an interesting discussion point: at what distance does the NFHT becomes indifferent to the local structures?  \\

Our approach uses the non-equilibrium Green's function (NEGF) formalism\cite{Walecka,Jauho,Mahan,Flensberg,Haug,Ashcroft} to describe the system's interaction which includes the self interactions (screening effects) and the energy transfer itself.  The latter has been excellently formulated by Caroli et al. \cite{Caroli}; this shall be referred as the Caroli formula.  Here, we shall present one method of deriving the Caroli formula based on Joule heating principle (another derivation using the “Poynting scalar" is presented in Appendix).  One possible extension to the Caroli formula is the energy transfer formula developed by Meir and Wingreen that takes into account electron tunneling between the objects \cite{Meir-Wingreen}.\\

Ultimately, our work is a continuation of recent attempts to determine a general rule of thumb on how a system's geometry and symmetry affect NFHT.  In particular, we aim to extend earlier works \cite{Jiang,Main} by enhancing computational efficiency using a reasonably simplified model as well as using it to analyze our two fundamental yet extensive geometries.

\section{Method}
	
\begin{figure}[h!]
\centering
\includegraphics[width=8cm]{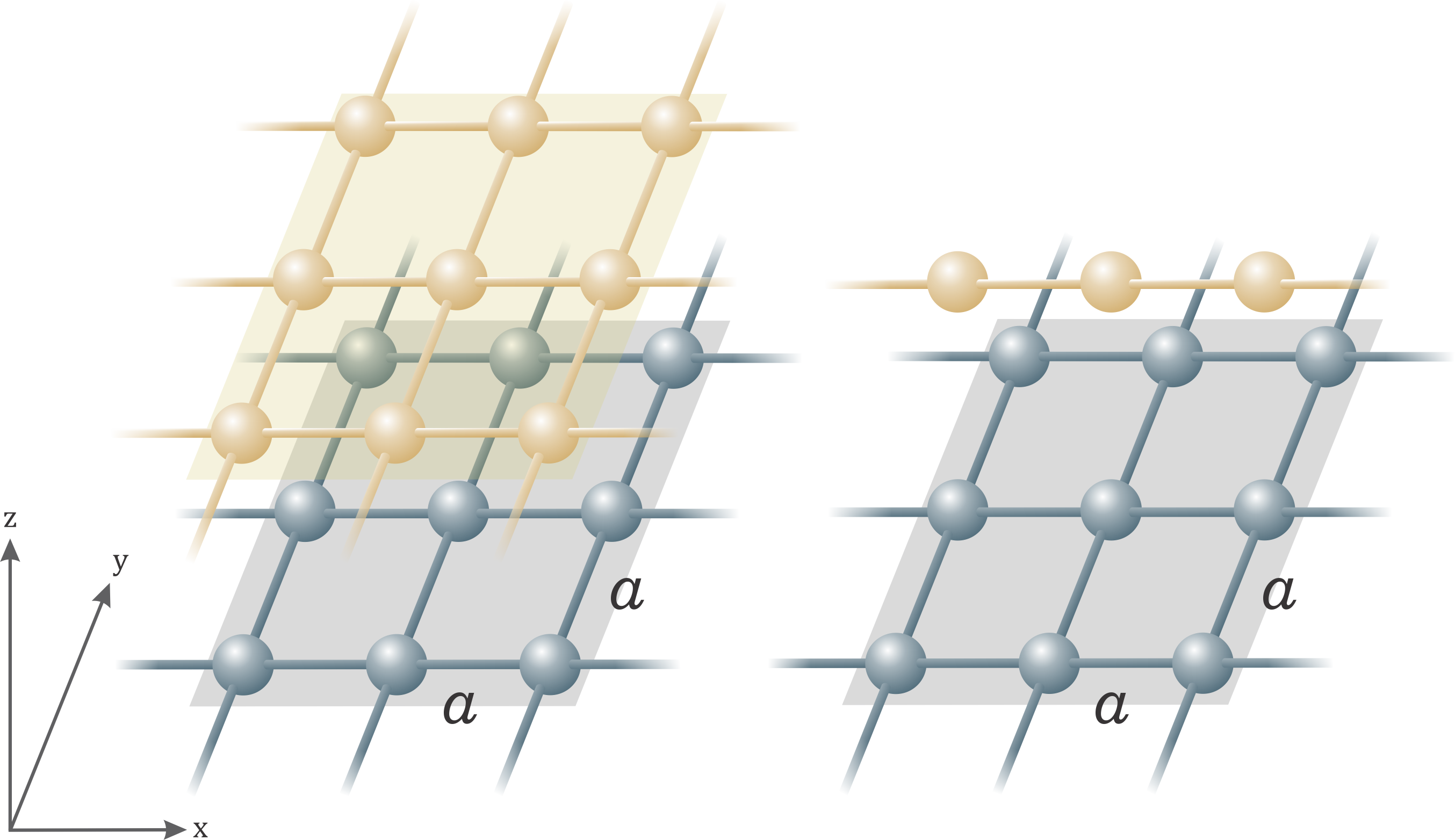}
\caption{(Left) Parallel surface system model. The graphene is an infinite
2D surface of simple lattice and the lattice points of the two surfaces are aligned.  (Right) Line-surface system model. The nanotube is an infinite 1D line lattice and its configuration is aligned in parallel with the surface.  All objects have equal lattice parameter $a$ in all directions.}
\end{figure}

The simplified system models for the graphene and nanotube are shown in Fig.~1.  In equilibrium, the electrons in both objects are governed by a tight binding Hamiltonian \cite{Wallace} which takes the form \cite{Netto}  $H=-t∑_{i,j}(c_i^{†} c_j+{\rm H.c.})$, describing electron hoppings with parameter $t$.  The $i,j$ summation is such that an electron can only jump to its nearest neighbours.  \\

The rationale behind this profoundly simplified structure and physical model includes a few lines of reasonings.  First, the Caroli formula for heat flux requires the knowledge of ${\mathrm \Pi}^{r}({\bf k},\omega)$, the self-energy.  ${\mathrm \Pi}^{r}({\bf k},\omega)$ describes the screening effect due to electron-electron interaction in an object and is related to the dielectric function $\epsilon$ by $\epsilon=1-v{\mathrm \Pi}^r$, $v$ being the Coulomb matrix.  Such quantity depends on the dispersion relation of the material that is deduced from its Hamiltonian.\\

Using the fundamental tight binding model, the dispersion relations take remarkably simple forms in the long-wave limit, namely linear for the surface and quadratic for the one-dimensional lattice.  In fact, a tight binding graphene is a unique two-dimensional system admitting such linear, Dirac cone dispersion \cite{Sarma}.  \\

The second reason is obvious: for the preservation of computational resources.  Our prediction is that such variation of local structure and the system's relative spatial and angular position shall have negligible impact on the heat flux when $d \gg a$; a conjecture we can confirm by comparing our final result to the earlier works.  \\

The last reason is that on top of linear dispersion, a tight binding graphene can be simplified from its double layer structure into a single one due to the fact that $t' \ll t$ ($t'$ being the hopping parameter in-between layers); a generally accurate statement as established using ab initio methods as well as actual experiments \cite{Reich,Deacon}.  Either way, any errors due to this simplification are diminished as the long wave limit is considered.  Hence in our model the electrons are confined in a single layer sublattice. \\

To obtain the self-energy formula we adopt the random phase approximation (RPA) \cite{RPA} scheme for the electrons interaction ${\mathrm{\Pi}}^r (t,0)=G_0^r (t) G_0^< (-t)+G_0^< (t)G_0^a (-t)$ which is then transformed to the $({\bf k},ω)$ representation ${\mathrm{\Pi}}^r ({\bf k},ω)=∑_l∫_{-∞}^{+∞}dt{\mathrm{\Pi}}_l^r (t,0) e^{-i\mathbf{k ⋅R_l}+iωt}$.  Note that $G_0$ denotes the known electron's equilibrium Green's function.  We then obtain the following expression for ${\mathrm \Pi}^r$ also known as the Linhard's function \cite{Linhard}\\

\begin{align}
{\mathrm{\Pi}}^r ({\bf k},ω)=-\frac{2e^2}{N_p} ∑_p\frac{f_{\bf p}-f_{{\bf p}-{\bf k}}}{ℏω+iη-ε_{\bf p}-ε_{{\bf p}-{\bf k}}}{\rm .}
\end{align}

The low energy limit dispersion relations $ε_{q}$, which for the two objects are given by $ε_{{\bf q}\:{ surface}}=\pm \frac{3at}{2}|\mathbf{q}|= \pm v_F ℏ|\mathbf{q}|$ and $ε_{q\:{ line}}=tq^2 a^2$ are to be substituted.  Temperature and chemical potential $\mu$ dependence are embedded in $f_{\bf p}$, the Fermi distribution.  We shall fix the other physical parameters for the above: $a=0.1$ nm, $t=2.7 \, {\rm eV}$, $η=0.0033 \, {\rm eV}$, and hence $v_F\approx97928 \, {\rm m/s}$.  \\

If a real hexagonal lattice model were used as in Ref.\cite{Jiang}, the same formula still applies save for an additional geometric phase matrix inside the summation.  Consideration of this term uses significant computation power and is the main cause for the arduous and lengthy computations, yet its impact on the main result is generally insignificant as we have argued.  \\

To preserve our computational resources further, we cite from earlier works expressions equivalent to the above expressions for zero temperature ${\mathrm \Pi}^r$ that gave an explicit expression without summation.  Here, we refer to Wunsch et al. \cite{Guinea} (another equivalent formulation was derived by  Hwang and Sarma \cite{Sarma}) for the surface ${\mathrm \Pi}^{r}$ and to Mihaila \cite{Mihaila} for the line's.  Note that Ref.\cite{Guinea} grants a formula that retains the aforementioned phase matrix factor; nevertheless, it converges to Eq.(1) with our local structure simplifications when $d\rightarrow \infty$.\\

One method of deriving the Caroli formula is via Joule heating, as suggested by Yu et al\cite{Yu}.  In the scheme of heat transfer by Coulomb interactions, the heat transferred from object 2 to object 1 is essentially the product of the change of induced charges in object 1 and the scalar potential at object 1 due to charge fluctuations at object 2.  More precisely, \\

\begin{align}
J_{2→1}=-\langle (\frac{dq_{1 i}}{dt})^T \psi_1 \rangle
\end{align}

where $q_{1 i}={\mathrm{\Pi}}_1^r \psi_1$  is the induced charge on object 1 after its screening takes effect and $\psi_1=D_{12}^r ξ_2$ is the scalar potential in object 1 due to charge fluctuations in object 2, $ξ_2$. These are formally expressed as \\

\begin{align*}
\psi_1 (t)&=∫dt'D_{12}^r (t-t')ξ_1 (t'){\rm ,}\\
q_{1 i} (t)&=∫dt'' ∫dt'{\mathrm{\Pi}}_1^r (t-t')D_{12}^r (t'-t'')ξ_1 (t''){\rm .}
\end{align*}

To avoid the multiple convolutions, we transfer the above expressions into the frequency domain.  Furthermore, the scalar photon Green's function relation\\ $-(\frac{i}{ℏ}) \langle \psi(t)\psi(t' )^T \rangle =  \frac{D^> (t-t')+D^< (t-t')}{2}$ together with the Keldysh equations \cite{Keldysh} $D^{<,>}=D^r {\mathrm \Pi}^{<,>} D^a$ are utilised to relate $ξ$ with $\bar{{\mathrm \Pi}}=\frac{{\mathrm \Pi}^>+{\mathrm \Pi}^<}{2}$.  \\

\begin{align*}
J_{2→1}&=∫_{-∞}^{+∞}\frac{dω}{2\pi} ℏω{\rm Tr} \bigl( D_{21}^a {\mathrm{\Pi}}_1^a D_{12}^r \overline{{\mathrm{\Pi}}}_2 \bigr) \nonumber\\
&=∫_{0}^{+∞}\frac{dω}{2\pi} ℏω(N_2+\frac{1}{2}){\rm Tr} \bigl( 4D_{21}^a {\rm Im}({\mathrm{\Pi}}_1^r ) D_{12}^r {\rm Im}({\mathrm{\Pi}}_2^r )\bigr){\rm .}
\end{align*}

Here, the condition of local equilibrium approximation (LEA) is applied to the ${\mathrm \Pi}$'s; this is a demand that they conform to the fluctuation dissipation theorem \cite{Jiang}: $\overline{{\mathrm{\Pi}}}=(N+\frac{1}{2})i({\mathrm{\Pi}}^r-{\mathrm{\Pi}}^a )=-(2N+1){\rm Im}({\mathrm{\Pi}}^r )$ where $N$ is the Bose distribution function.\\

Notice that it is unreasonable to expect the above LEA and fluctuation dissipation theorem to hold at exceedingly short distances.  In practice, connection to thermal bath is required to maintain local equilibrium of an object.  The heat flux result will however be affected by the exact manner in which we connect these baths; this would in turn defeat the purpose of our geometric effect on NFHT study. Fortunately, our geometric study concerns semi-infinite models, in which we can - in theory - connect the thermal baths at infinity. This will ensure local equilibrium in each object, yet not altering the results with the technicalities.  \\

In any case, we will always avoid using our model on extremely small distances, i.e., the condition in which LEA might fail will always be avoided.  This is - as we will demonstrate and discuss in the following sections - due to the existence of a critical distance of which our model must exceed to maintain its validity.  To put it shortly, on top of LEA, our simplified band structure/dispersion model and cited self-energy expressions compatibility with Eq.(1) might also fail below the critical distance.  Not to mention, at distances about 1 to 2 {\AA}, the electrons might begin to jump from one object to the other; this by itself limits the validity of our no-tunneling assumption to distances of about 0.5 nm or larger.\\

Repeating the calculations for $J_{1→2}$ and subtracting from the above yields the Caroli formula\\

\begin{align}
J=∫_{0}^{+∞}\frac{dω}{2 \pi} ℏω(N_1-N_2 ){\rm Tr} \bigl( 4D_{21}^r {\rm Im}({\mathrm{\Pi}}_1^r ) D_{12}^a {\rm Im}({\mathrm{\Pi}}_2^r ) \bigr){\rm .}
\end{align}

The scalar photon's Green's function $D^r$ follows the Dyson equation\\

\begin{align}
D^r=D_0^r+D_0^r {\mathrm{\Pi}}^r D^r{\rm .}
\end{align}

In this case the scalar photon's $D_0^r$ is simply the instantaneous Coulomb interaction\\

\begin{align}
D_0^r (\mathbf{r},\mathbf{r'},\omega)=\frac{1}{4\pi ε_0 |\mathbf{r}-\mathbf{r'} |}{\rm .}
\end{align}

As our two systems have fundamentally different symmetries, different representations for $D_0^r$ are used.  For the parallel surface,\\

\begin{equation}
D_0 (k_x,k_y,\omega,z,z' )=\frac{i}{2\tilde{k}ε_0 a^2 }
\begin{bmatrix}
    1     & e^{i\tilde{k}d}\\
    e^{i\tilde{k}d}       & 1
\end{bmatrix}{\rm ,}\\
\end{equation}

where $\tilde{k}$ represents $\sqrt{(i\delta)^2-(k_x^2+k_y^2)}$, $\delta$ being a small regularizing parameter which we set to $0.0033 \frac{\mu}{\hbar v_F}$.  For the line-surface on the other hand,\\

\begin{equation}
D_0^r (k_x,\omega,y,y',z,z')=\frac{1}{4\pi ε_0a}\Lambda{\rm ,}
\end{equation}

\begin{equation}
\Lambda=
\begin{bmatrix}
    \sum_{n\neq0} \frac{e^{-ik_x na}}{n}     & 2K_0 (\sqrt{m^2+(d/a)^2 } ak_x )\\
    2K_0 (\sqrt{m^2+(d/a)^2 } ak_x )       & 2K_0 (|m-m' | ak_x )
\end{bmatrix}{\rm .}\nonumber\\
\end{equation}

In the above $K_0$ denotes the modified Bessel function of the second kind of order 0. This function is merely the result of the single direction Fourier transform of Eq.(5); the definition $K_0 (\alpha)=\frac{1}{2} ∫_{-∞}^{+∞} dt \frac{e^{i \alpha t}}{\sqrt{1+t^2} }$ is used.  Also note that the first entry of $\Lambda$ is understood to be the result of FFT, i.e., $n\rightarrow \infty$.  Finally, $m$ and $m'$ are numbers indexing $y$ and $y'$, respectively.

\section{Results and Discussion}

\subsection{Parallel Surface Geometry}

For the sake of comparison with Ref.\cite{Jiang}, we chose our parameters to match those in their work.  Fig.~2 compares the two results - represented by the $J/J_{BB}$ vs $d$ curves, $J_{BB}$ being the heat current density corresponding to standard black body radiation - head on.\\

\begin{figure}[h!]
\centering
\includegraphics[width=8cm]{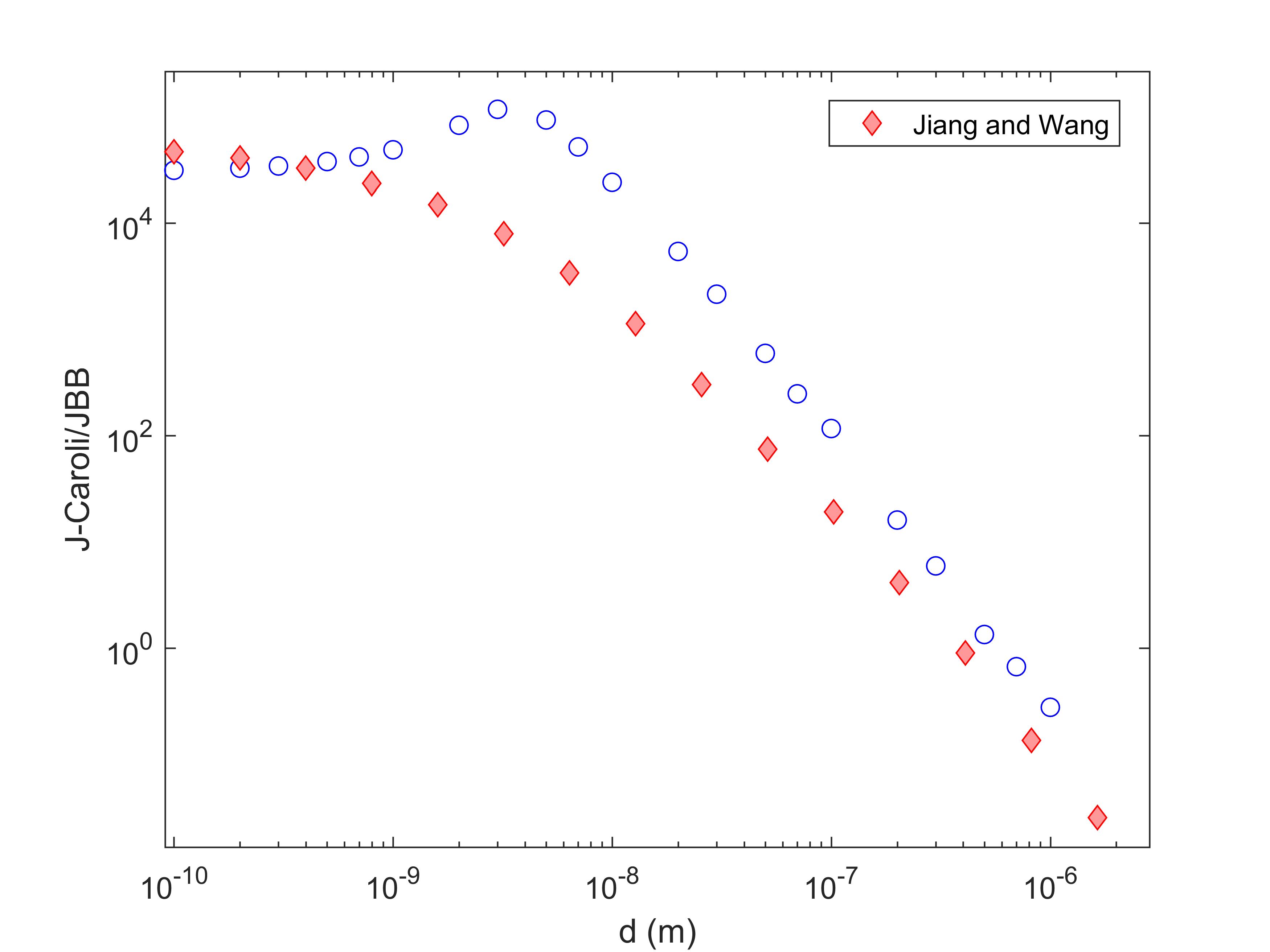}
\caption{Heat transfer enhancement vs distance for parallel surface system compared with Ref.\cite{Jiang}.  $μ_1=μ_2=0.1 \, {\rm eV}$, $T_1=300$ K, $T_2=1000$ K.}
\end{figure}

\begin{figure}[h!]
\centering
\includegraphics[width=8cm]{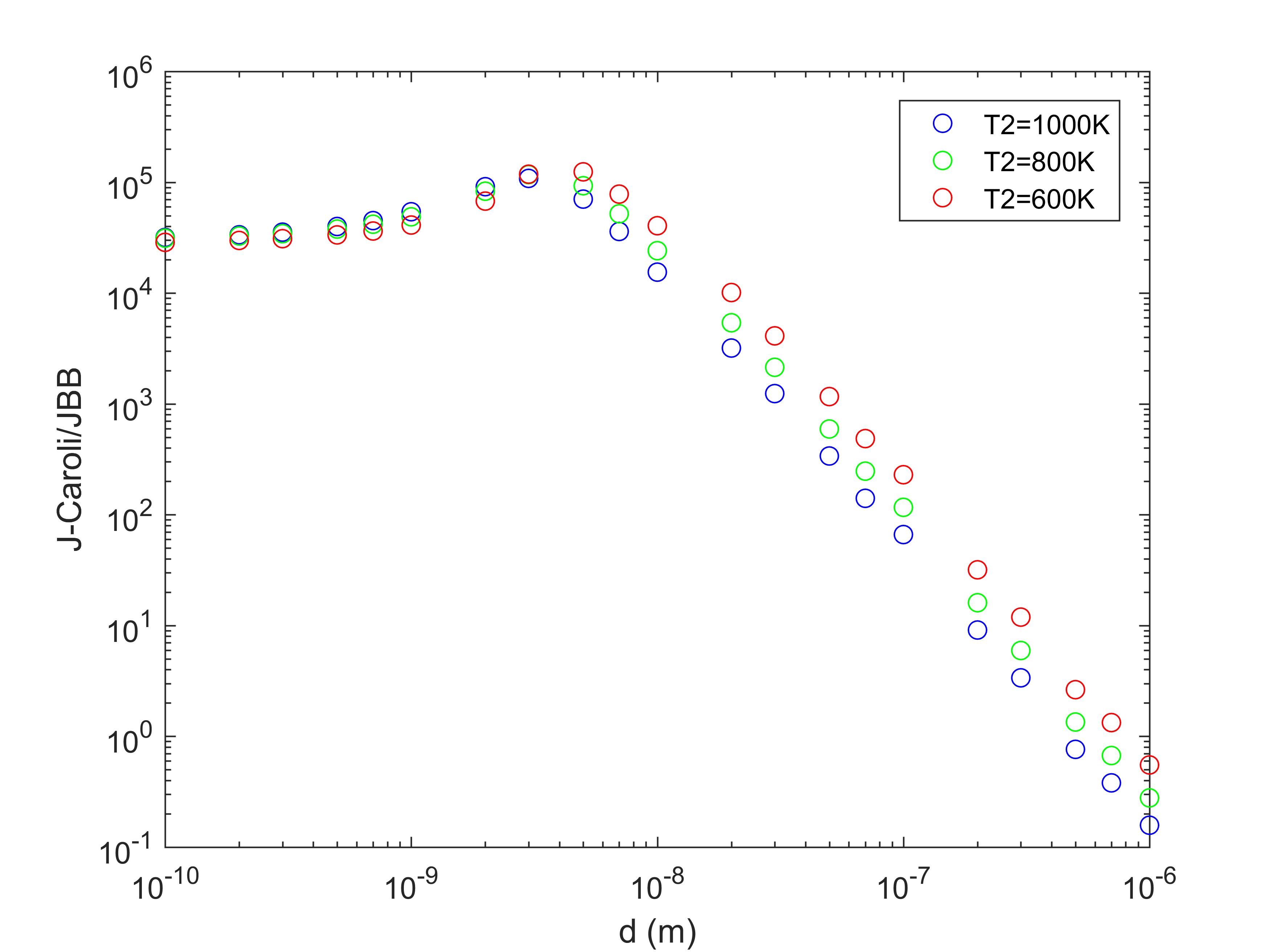}
\caption{Heat transfer enhancement vs distance for parallel surface system under temperature variation.  $μ_1=μ_2=0.1 \, {\rm eV}$, $T_1=300$ K.}
\end{figure}

Two observations are immediate: the two results converge in slope and value from $d\approx100$ nm and they differ most significantly at $d\approx5$ nm where a somewhat unexpected peak arises in our result.  The first observation is expected: we have reproduced the same conclusion as Ref.\cite{Jiang} that for two parallel surface system - irrespective of the local geometry - heat transfer is asymptotically scaled as $J \sim d^{-2.2}$.  Furthermore, we found such relation to be invariant under variations of physical parameters $t$ and $a$ (and as we shall discuss, to temperature and chemical potential as well), save for distance rescalings.\\

The peak on the other hand represents the main discrepancy between the two models, induced most probably by the integrated failures of our LEA, simplified dispersion model and self-energy expressions as we predicted.  Hence, a valuable interpretation is that this peak represents the critical distance above which our model accurately replicates a real lattice model graphene; indeed, the “correct" linear regime emergence immediately follows this peak.  \\

Another unforeseen outcome is that both plots converge once again in the limit $d \xrightarrow{} 0$, contrary to our conjecture that the two models' fundamentally different structures at local scale would generally imply significant discrepancies on small distances.  Note that by no means this generally guarantees our model's accuracy at very small distance; afterall, it is only formulated to predict heat transfer at large distances as discussed.    \\

Following Ref.\cite{Jiang}, the temperature variation is also analyzed, this is shown in Fig.~3.  Similar conclusions are drawn: remarkable agreement are achieved on the two extreme ends while the peak anomaly is observed in the neighbourhood of $d\approx5$ nm.  On the other hand, the effect of increasing temperature itself appears to diminish the heat transfer by a constant (logarithmic) value throughout the linear regime; it however does not affect the curve's slope.  Note that this adverse relationship implies that the Coulomb force NFHT is less sensitive to temperature than that of the black body radiation.\\

In our model, variation of doping levels, i.e., chemical potentials $\mu$ are trickier to apply since we need to ensure the validity of the Dirac cone dispersion that was assumed.  No accurate comparison with Ref.\cite{Jiang} is feasible in this case as exceedingly high values of $\mu$ are required.  Nevertheless, under a fairly wide range of $\mu$, the $-2.2$ exponent is still preserved at large distances, establishing itself as an invariant for a given geometry.

\subsection{Line-Surface Geometry}

Physical parameters used in this section follows the previous ones with the exception of the line's $\mu$ which we set to 0.05 eV to avoid violating the necessary condition of $\mu/t \ll 1$ that ensures the validity of the quadratic dispersion.  Also note that the system's dimensionality necessitates $J$ in this section to be heat current per unit length instead of area; comparison with $J_{BB}$ is thus not practical.\\

Despite their ostensibly similar profile, the heat transfer curve for the line-surface geometry (Fig.~4) has a fundamentally different quality when compared to its parallel surface counterpart.  In this system, the linear regime encompasses a wider region and emerges from a distance less than $1$ nm.  Furthermore, the slope of the heat transfer curve is much steeper, approximately $-5.1$.  This comparison suggests that the enhancement effect of NFHT is more pronounced and global in this geometry.  Instabilities on the other hand emerge when the distance falls below $d\approx0.9$ nm; these presumably signify the breakdown of our model and $d=0.9$ nm plays the role of the “critical distance" in this system.   \\

\begin{figure}[h!]
\centering
\includegraphics[width=8cm]{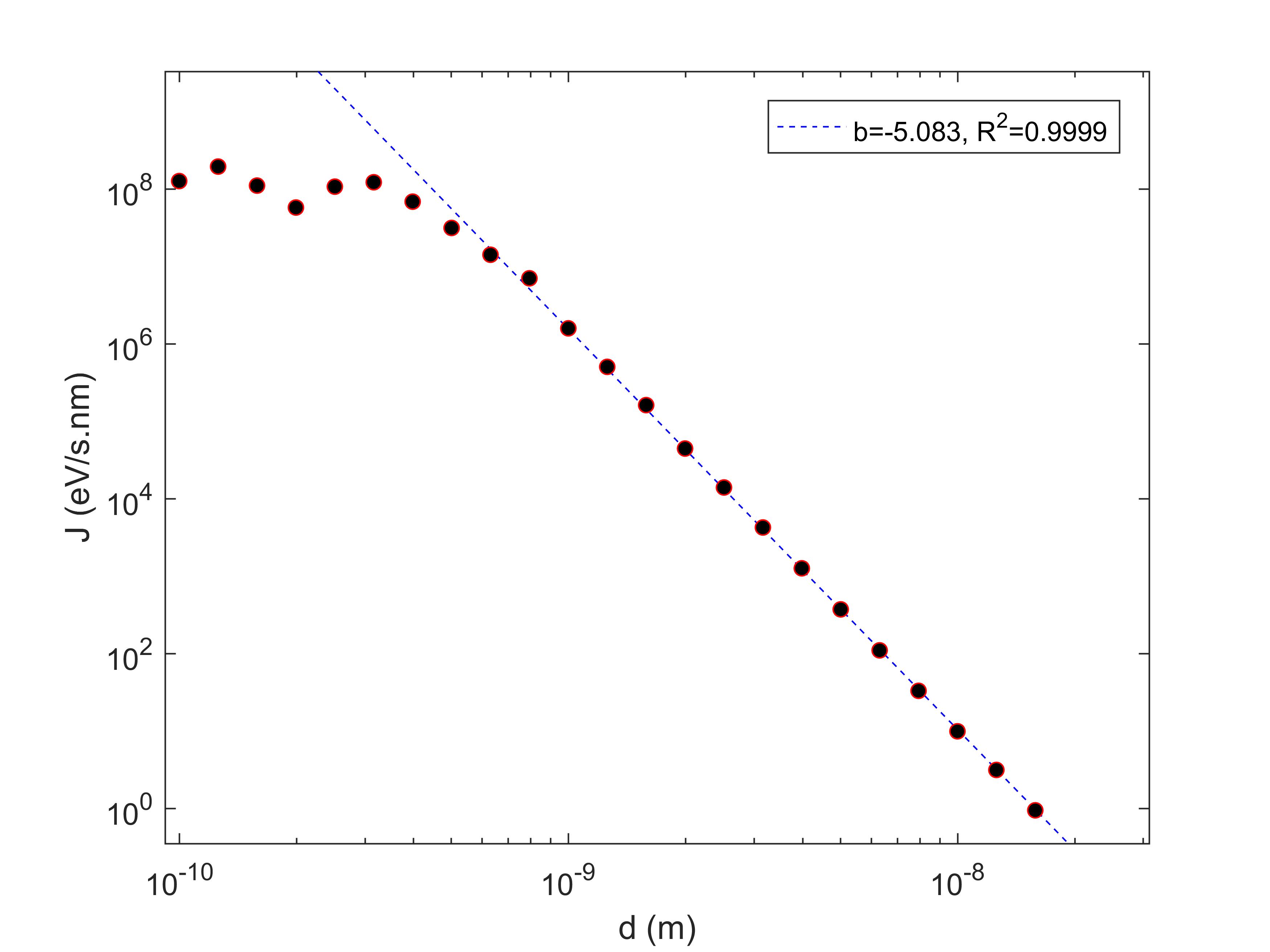}
\caption{Heat transfer vs distance for line-surface system with a linear fit on its linear regime.  $μ_{line}=0.05 {\rm eV}$, $μ_{surface}=0.1 {\rm eV}$, $T_1=300$ K, $T_2=1000$ K, $N_y=71$.}
\end{figure}

\begin{figure}[h!]
\centering
\includegraphics[width=8cm]{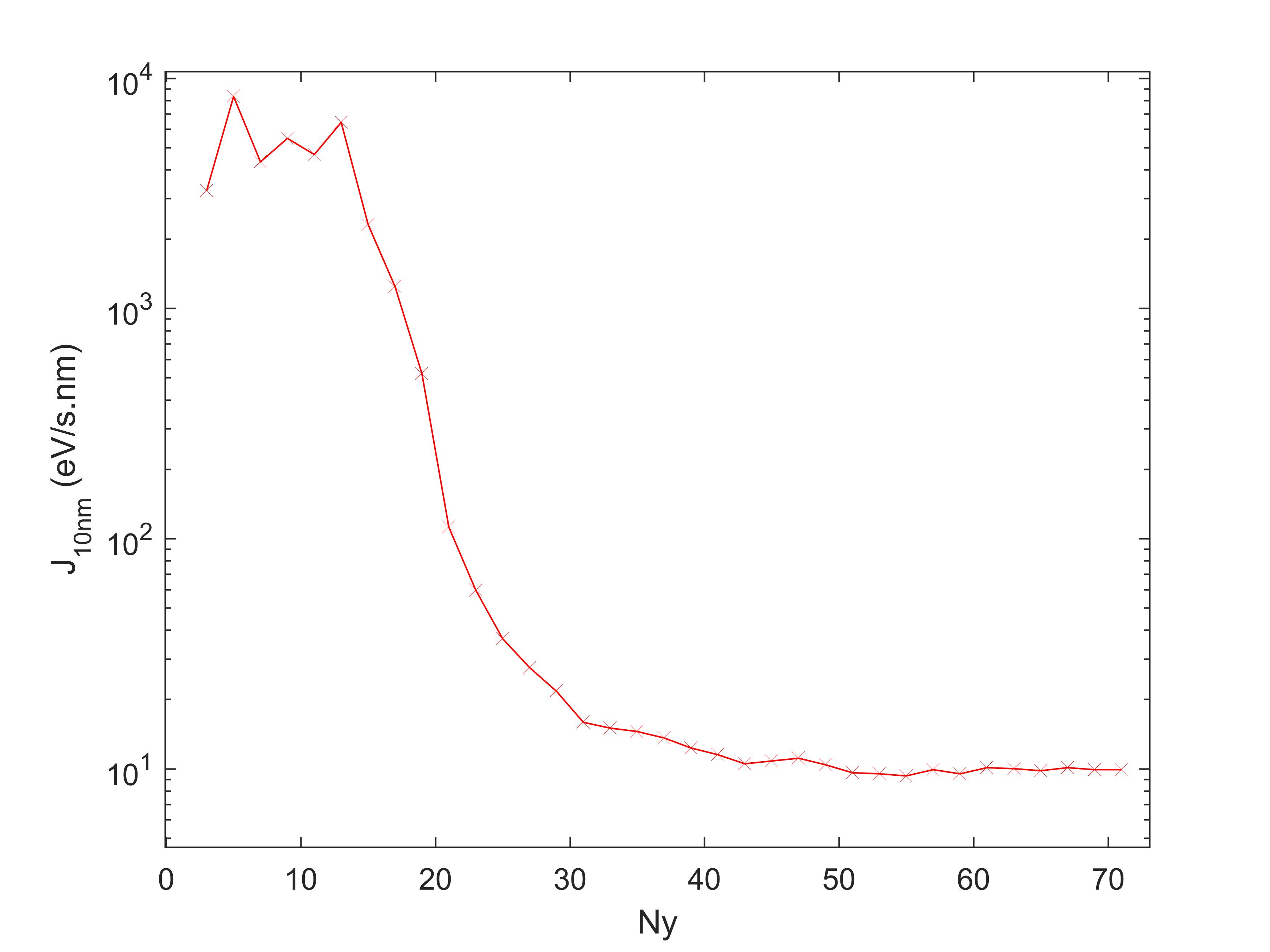}
\caption{Convergence of $J$ at $10$ nm under $N_y$ variation.  Other parameters are as in Fig.~4.}
\end{figure}

\begin{figure}[h!]
\centering
\includegraphics[width=8cm]{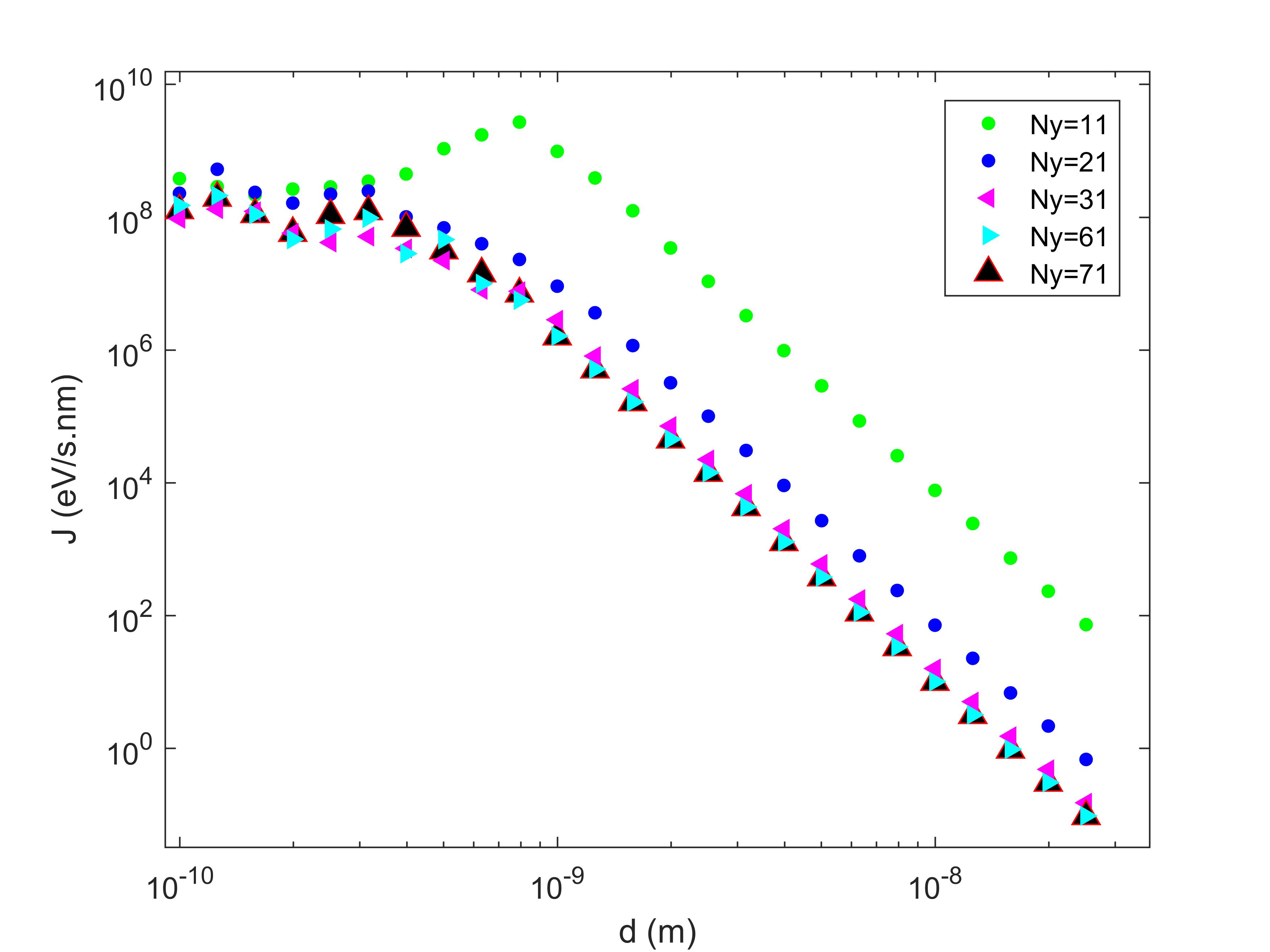}
\caption{Heat transfer curves under various $N_y$.  Other parameters are as in Fig.~4.}
\end{figure}

Note that unlike in the direction $x$ parallel to the line, in the transverse direction $y$, a reasonably small value for the number of lattice points $N_y$ is desired.  This is due to the fact that while $N_x$ corresponds only to the array size on the FFT processes, $N_y$ also corresponds to the characteristic matrices dimension and as such would largely impact the computational arduousness.\\

The convergence of $J$ under $N_y$ - a study of size effect on NFHT - is closely examined in Fig.~5.  This behavior is typical especially in the linear regime of the curves.  It is then sensible to choose for example $N_y=71$ as in Fig.~4: a value demonstrated to yield accurate results, yet reasonably small to maximize computational efficiency.  \\

To complement Fig.~5, Fig.~6 presents the heat transfer curves for various $N_y$.  It must be noted however, that plugging too small of a number for $N_y$ will introduce errors in our computations such as in the FFT processes; one cannot plug $N_y=1$ and expect an accurate result for a limiting case of line vs line NFHT.  Nonetheless, Fig.~6 shows that the curves' slope and profile exhibit negligible variation when $N_y$ is varied,  the curves merely shift vertically and converge approximately after $N_y\approx31$.  This behavior is intriguing since we expected more significant discrepancies of the slopes or general profile for objects with different sizes.\\

Lastly, variations of thermodynamic and other physical parameters introduce no interesting or new behaviors on these curves.  Under $T$ variation, the linear regimes of the curves merely shift vertically similar to the observed behavior in the parallel surface geometry.  This is expected since the $T$ dependent functions are essentially the same in both systems.  This mundane kind of variation is also observed under $\mu$ variation, no matter which object is subjected to the change.  Again, this is similar to the typical behavior found in the parallel surface.

\subsection{NFHT Spectral Analysis}

Even though a complete description of the surface modes that contribute to NFHT are nearly impossible to be obtained at this point, its spectral analysis e.g. transmission coefficient ${\rm Tr} \bigl( 4 D_{21}^a {\rm Im}({\mathrm{\Pi}}_1^r ) D_{12}^r {\rm Im}({\mathrm{\Pi}}_2^r )\bigr)$ as function of frequency $\omega$ is available to help our understanding of the involved mechanism.  From Fig. 7 and Fig. 8 the dominating frequencies of NFHT can be deduced.\\

\begin{figure}[h!]
\centering
\includegraphics[width=7cm]{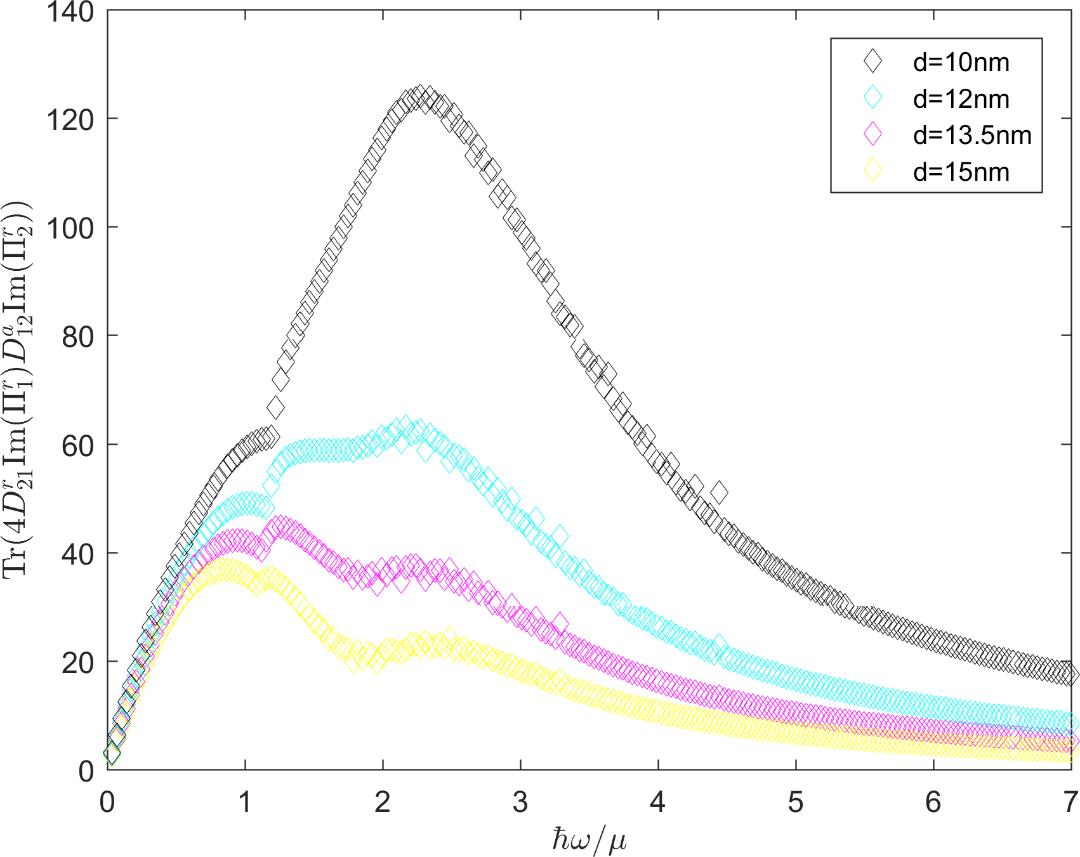}
\includegraphics[width=7cm]{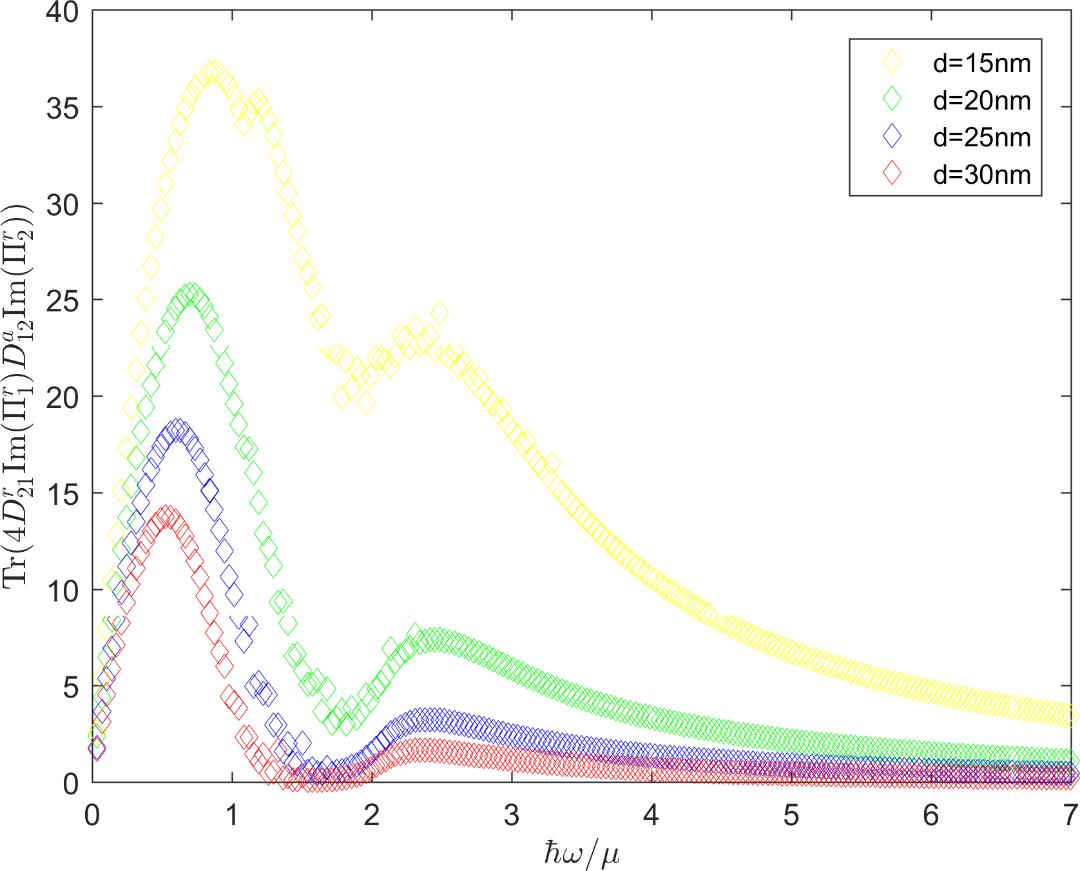}
\caption{The transmission coefficient under varying frequency represented by the dimensionless $\frac{\hbar \omega}{\mu}$ in the parallel surface system.  Parameters are as in Fig.~4.}
\end{figure}

\begin{figure}[h!]
\centering
\includegraphics[width=7cm]{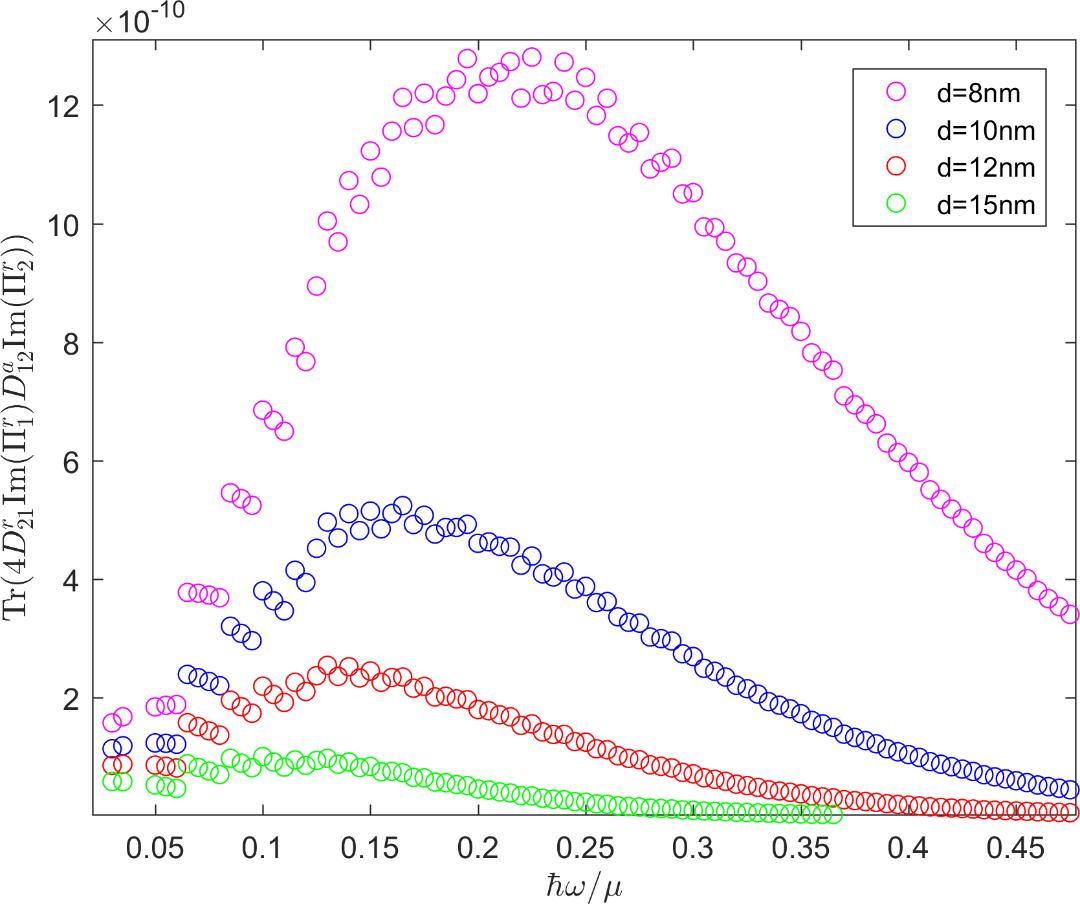}
\caption{The transmission coefficient under varying frequency in the line-surface system.  Parameters are as in Fig.~4.}
\end{figure}

In the parallel surface system, two peaks in the spectral curve are generally observed.  It is interesting to note that the primary lower frequency peak dominates when $d$ is greater than apporximately 12 nm whereas the secondary higher frequency peak dominates below this distance.  In the limiting case $d\rightarrow \infty$, the low frequency mode whose value of $\omega$ is slowly decreasing with increasing distance becomes the singular mode of the NFHT; its asymptotic value is approximately $7.93\times10^{13}\text{ }{\rm s}^{-1}$. Similarly, the results suggest the high frequency mode - whose value is rather consistent at $\omega \approx 3.49\times10^{14}\text{ }{\rm s}^{-1}$ - to be the singular mode at small distances; this mode generally has wider breadth than the former. \\

In the line-surface system, the NFHT is generally dominated by a single frequency mode.  At $d=8$ nm, the value of this frequency is $\omega=1.60\times 10^{13}\text{ }{\rm s}^{-1}$ and it decreases more rapidly - compared to the high frequency mode of the parallel surface system - with no apparent asymptotic value as $d$ increases.  The height of the peaks likewise decays more rapidly with increasing distance in this case as expected from its steeper heat transfer curve.

\section{Conclusion}

With our proposed simplified models of graphene and nanotube, we computationally simulated and analyzed the Coulomb force mediated NFHT for two types of geometry.  We found these highly efficient models capable of producing accurate results for a reasonably wide range of physical conditions albeit raising some anomalies typically at very small object to object distances.  This relative success of our model is significant for the study of NFHT: in future works where a large number of data is required or a more complex geometry is to be considered, its simplicity will prove	 useful.  \\

For the parallel surface system we found the heat transfer dependence on distance to be $\sim d^{-2.2}$ in exact agreement with the conclusion of Ref.\cite{Jiang}.  Second, we found that in such geometry, the “critical distance" that separates the accurate and inaccurate region to be approximately $5$ nm.  Third, applying the same procedure on the line-surface geometry we concluded that $J\sim d^{-5.1}$; this was stemmed from the extensive linear regime of the curve that unexpectedly begins from distances less than 1 nm, challenging our supposition that our model shall fail at distances of the order of $a$.  Erratic behavior and in all probability inaccurate results begins below a much smaller “critical distance" of $d\approx0.9$ nm.  Fourth, we found that for both geometries physical parameters such as chemical potential and temperature and even the lattice number $N_y$ do not change the profile (i.e. the slopes of the curves) and their impact is merely a constant shift in the linear region.  Lastly, we studied the transmission spectrum of NFHT in the two systems and found out the many characteristic discrepancies in their respective dominating heat transfer mode.  \\

Future works following this study may include the use of our model to study NFHT where electric currents are allowed on the objects \cite{Jiebin2}.  Theoretically, such systems will exhibit heat transfer even when both objects are completely identical in terms of thermodynamic parameters ($μ$ and $T$).  Modifications are required on the Caroli formula and formulas for the self-energies.  More precisely, the ω dependent functions must now take into account the Doppler shifts due to the current density.

\begin{appendices}

\setcounter{figure}{0} \renewcommand{\thefigure}{B.\arabic{figure}}
\setcounter{secnumdepth}{0}
\section{Appendix: Caroli Formula Derivation Using Poynting Scalar Method}
We presented here a brief note on the Poynting scalar approach \cite{Jiebin} for the computation of heat transfer.  We began by considering the electromagnetic energy density formula in terms of scalar field and the $c\rightarrow \infty$ limit of the Poisson equation \\

\begin{align*}
U=-\frac{1}{2} ε_0 \Bigl( \frac{\dot{φ}^2}{c^2}+(∇φ)^2 \Bigr) {\rm ,}
\end{align*}

\begin{align*}
\frac{1}{c^2}  \ddot{φ}(\mathbf{r},t)-∇^2 φ(\mathbf{r},t)=0 {\rm .}
\end{align*}

Note that $φ$ by definition is related to the photon's greater Green's function by \\

\begin{align*}
D^> (\mathbf{r},t,\mathbf{r'},t' )=-\frac{i}{ℏ} \langle φ(\mathbf{r},t),φ(\mathbf{r'},t' ) \rangle {\rm .}
\end{align*}

The divergence of heat transfer density $\mathbf{J}$ is given by the intrinsic decrease rate of energy density $-\frac{∂U}{∂t}$; this is just the usual statement of continuity,\\

\begin{align*}
∇\boldsymbol{\cdot}\mathbf{J}&=-\frac{∂U}{∂t}=\frac{1}{2} ε_0 [\frac{2\dot{φ}\ddot{φ}}{c^2} +2∇φ\boldsymbol{\cdot}∇\dot{φ}]\nonumber\\
&=ε_0 [\dot{φ}∇^2 φ+∇φ\boldsymbol{\cdot}∇\dot{φ} ]\nonumber\\
&=ε_0 ∇\boldsymbol{\cdot}[\dot{φ}∇φ] {\rm .}
\end{align*}

Hence, the formula of energy transfer density in terms of the scalar field (and its derivatives) is\\

\begin{align*}
\mathbf{J}(\mathbf{r},t)=ε_0 \dot{φ}(\mathbf{r},t)∇φ(\mathbf{r},t) {\rm .}
\end{align*}

The planar symmetry of the system concerned in this paper guarantees that $\mathbf{J}$ will only be a function of $z$, while the steady state requirement demands it not to depend on time as well.  $\mathbf{J}(\mathbf{r},t)$ is thus simplified to $J(z)$.  We then consider $J$ as a fluctuating quantity\\

\begin{align*}
\langle J(z) \rangle=ε_0  \frac{∂}{∂t}  \frac{∂}{∂z'} \langle φ(z,t)φ(z',0) \rangle |_{t=0.z'=z}
\end{align*}

where we have utilized the time translational invariance to set the second time argument $t'=0$.  Substituting the bracket by $D^>$ yields\\

\begin{align*}
\langle J(z) \rangle=ε_0  \frac{∂}{∂t}  \frac{∂}{∂z'}(iℏ)D^> (\mathbf{r},\mathbf{r'},t,0) |_{t=0,z'=z} {\rm .}
\end{align*}

To avoid the troublesome $\frac{∂}{∂t}$, we transform the above into $\omega$ representation\\

\begin{align*}
\langle J_z (z) \rangle&=ε_0 ∫_{-∞}^{+∞}\frac{dω}{2\pi}  \frac{∂}{∂t}  \frac{∂}{∂z'} e^{-iωt} (iℏ)D^> (\mathbf{r},\mathbf{r'}ω) |_{t=0,z'=z}\nonumber\\
&=ε_0 ∫_{-∞}^{+∞}\frac{(ℏω)dω}{2\pi}  \frac{∂}{∂z'} D^> (\mathbf{r},\mathbf{r'}ω) |_{z'=z}{\rm .}
\end{align*}

The equivalence of this formula with Eq.(3) has been proven for long wave limit by Ref.\cite{Jiang}, and can be shown for the general case using direct computation.

\end{appendices}

\bibliographystyle{apsrev4-2}

\end{document}